\documentclass[ twocolumn,08 pt, amsmath,amssymb, aps,fleqn]{revtex4-1}
\usepackage[hidelinks]{hyperref}
\usepackage[english]{babel}
\usepackage{mathtools}
\usepackage{tabularx}
\usepackage{multirow}
\usepackage{comment}
\usepackage{graphicx}
\usepackage{color}
\usepackage{soul,cancel}
\begin{document}
\preprint{APS/123-QED}
\title{Conductance of inhomogeneous Luttinger liquids with a finite bandwidth}
\author{Joy Prakash Das} \author{Girish S. Setlur}\email{gsetlur@iitg.ernet.in}
\affiliation{Department of Physics \\ Indian Institute of Technology  Guwahati \\ Guwahati, Assam 781039, India}
%------------------------------------ABSTRACT now-------------------------------------------------------------------------------------------------------
\begin{abstract}
The finite-bandwidth conductance of a Luttinger liquid (LL) with a cluster of impurities is studied and its variation with respect to temperature is shown. The calculations are done using the correlation functions obtained using the powerful non-chiral bosonization technique (NCBT) . The results are compared with those obtained by Matveev, Yue and Glazman [K. Matveev et al., Phys. Rev. Lett. 71, 3351 (1993)] who deal with a weakly interacting LL.  By contrast, NCBT correctly provides the conductance for all values of the interaction strength (as well as the sign). In addition to finding perfect agreement with the results of Matveev et al. for both weakly repulsive and weakly attractive mutual interactions, we are also able to probe novel physics seen when the repulsion is strong - in the form of a weakly temperature dependent conductance when there is a definite relationship between the transmission amplitude of the non-interacting system and the holon velocity.  Secondly, an unusual high conductance for strongly repulsive mutual interactions is observed for a weak barrier at low temperatures. Lastly, inclusion of backward scattering leads to the non-monotonic temperature dependence of conductance when dealing with fermions with spin. This work is also important as a validation of the NCBT itself.

\end{abstract}

\maketitle
%---------------------------------------INTRODUCTION----------------------------------------------------------------------------
\section{Introduction}
It is well-known that the electrons in a clean one dimensional system move ballistically, with a quantized conductance \cite{beenakker1991solid}. However this motion  is heavily damped by the presence of even a small scatterer, which can be as drastic as `breaking the chain' for repulsive interactions \cite{kane1992transport}. In the absence of mutual interaction between the fermions, the conductance of such systems is simply related to the transmission coefficient by Landauer's formula \cite{landauer1957r}.  Inclusion of mutual interactions, which leads to a state described by the Luttinger liquid \cite{haldane1981luttinger}, gives rise to some interesting many-body physics which are quite different from those in higher dimensions. With the advent of technology, such systems are increasingly becoming physically realizable such as carbon nanotube \cite{bockrath1999luttinger, bockrath2001resonant, yao2000high}, semiconducting quantum wires \cite{auslaender2000experimental, yacoby1997magneto}, etc. which holds a promising future in terms of technology. Hence the study of transport properties of such systems is one of the main themes of research in condensed matter systems. For homogeneous systems, these systems are easily studied using bosonization methods while the introduction of impurities requires additional methods such as renormalization group, etc. \cite{giamarchi2004quantum, kane1992transport, meden2008fermionic} 

Transport and other physical properties of quantum systems can be studied if one is in possession of closed analytical expressions of the Green functions of such systems. To this end, a recently developed technique called non-chiral bosonization is employed here that yields the most singular part of the asymptotic Green functions of strongly inhomogeneous Luttinger liquids with arbitrary strengths of impurities as well as that of interactions \cite{das2018quantum}. The same method also yields the Green functions of a one step fermionic ladder system comprising two Luttinger liquids placed parallel to each other with a finite probability of hopping of electrons between a pair of opposing points \cite{das2017one}. 
The conductance of a clean Luttinger liquid is generally given by $g\hspace{0.05cm}e^2/h$ where `$g$' is the Luttinger parameter which is a function of the strength of mutual interactions \cite{kane1992transport, apel1982combined, ogata1994collapse}. But no renormalization of the universal conductance is required provided the electrons have a free behavior in the source and drain
reservoirs \cite{ponomarenko1995renormalization, maslov1995landauer}. However for weakly disordered quantum wires connected to non-interacting leads the conductance does scale with temperature and length of the wire \cite{maslov1995transport} which explains quasiballistic nature of electrons in GaAs quantum wires \cite{tarucha1995s}.  Kane and Fisher expressed conductance across a truly insulating link as a power law in temperature with an exponent that depends on the Luttinger parameter `$g$' \cite{kane1992transport}.   Matveev et al. \cite{matveev1993tunneling} used a simple renormalization group method to calculate the conductance of a weakly interacting electrons in 1D in presence of a scatterer of arbitrary strengths. They were able to describe the temperature dependence of conductance at any temperature, both for finite and infinite bandwidth.  Matveev  also studied \cite{matveev2004conductance} the effect of interactions on the conductance of a Luttinger liquid connecting two bulk leads. Ogata and Anderson \cite{ogata1993transport} studied conductivity of a Luttinger liquid using Green's functions and showed that if the spin-charge separation is taken into account, the resistivity has a linear temperature dependence. 
Exact conductance through point contacts in a Luttinger liquid is obtained by Fendley et al. \cite{fendley1995exact, fendley1995exact2}. More recently conductance has been studied using numerical methods like Monte Carlo simulations \cite{hamamoto2008numerical, morath2016conductance} and quantum simulations \cite{anthore2018circuit}.  
Aseev et al. \cite{aseev2018conductance} recently studied how the combined effect of multi-electron interaction and applied magnetic field leads to a gap in the spectrum, which in turn affects the temperature dependence of fractional conductance of a quantum wire. Other works on transport properties in 1D systems include study of  long range disorder  \cite{levchenko2010transport}, short range disorder \cite{furusaki1993furusaki, giamarchi1988t}, thermal transport \cite{kane1996thermal, krive1998thermal, degottardi2015electrical}, spin dependent transport \cite{balents2001spin}, frequency dependent transport \cite{ponomarenko1996frequency} and so on.

In this work, conductance is studied as a tunneling phenomenon in a Luttinger liquid with a cluster of impurities using the correlation functions obtained using NCBT. The next section describes the system that is studied, followed by a section where NCBT is briefly discussed. The subsequent sections presents the results of conductance calculations for the given class of systems and a detailed comparison with the work of Matveev et al. \cite{matveev1993tunneling} is made. While the latter is restricted to weak interactions, this work is able to probe novel physics seen for strong interactions. The temperature dependence of conductance is also favorably compared to numerical results. This work is also important as a validation of the NCBT itself.  However, this is somewhat superfluous as NCBT has been fully validated in earlier works \cite{das2018non} where it is shown that the Green functions obtained from this method obey the exact Schwinger Dyson equation and that they are also consistent with conventional perturbation theory. Also the exact tunneling density of states near the impurity for Luttinger parameter $K=1/2$ is reproduced using NCBT in an earlier work \cite{das2018friedel}. Published works that use NCBT include application to the study of a cluster of static impurities in a Luttinger liquid \cite{das2018quantum}, the one-step fermionic ladder \cite{das2017one} and the mobility of heavy particles in a Luttinger liquid \cite{das2018ponderous}.

%----------------------------SYSTEM DESCRIPTION-----------------------------------------------
\section{System description}

A Luttinger liquid with a cluster of impurities near the origin and short range forward scattering mutual interactions between the fermions is considered. The generic Hamiltonian of the system can be written as follows.
\small
\begin{equation}
\begin{aligned}
H =& \int^{\infty}_{-\infty} dx \mbox{    } \psi^{\dagger}(x) \left( - \frac{1}{2m} \partial_x^2 + V(x) \right) \psi(x)\\
  & \hspace{1cm} + \frac{1}{2} \int^{ \infty}_{-\infty} dx \int^{\infty}_{-\infty} dx^{'} \mbox{  }v(x-x^{'}) \mbox{   }
 \rho(x) \rho(x^{'})
\label{Hamiltonian}
\end{aligned}
\end{equation}
\normalsize
The first two terms are the kinetic energy and the potential energy terms, the latter representing the impurity cluster which is modeled as a finite sequence of barriers and wells around a point (taken to be the origin, $x=0$). The potential cluster, which breaks the homogeneity of the system, can be as simple as a delta impurity $V_0\delta(x)$, two delta impurities placed close to each other $V_0( \delta(x+a)+\delta(x-a))$, finite barrier/well $\pm V \theta(x+a)\theta(a-x)$ and so on, where $\theta(x)$ is the Heaviside step function. 
The third term represents the forward scattering mutual interaction term such that
\begin{equation}
\hspace{2 cm} v(x-x^{'}) = \frac{1}{L} \sum_{q}  v_q \mbox{ }e^{ -i q(x-x^{'}) } 
\label{vq}
\end{equation}

where $ v_q = 0 $ if $ |q| > \Lambda $ for some fixed bandwidth $ \Lambda \ll k_F $ and $ v_q = v_0 $ is a constant, otherwise. 
The goal of this work is to calculate the tunneling conductance of these systems using the correlation functions obtained in an earlier work \cite{das2018quantum}. For an analytical solution to be feasible, the RPA (random phase approximation) is imposed on the system. In this limit, both the Fermi momentum and the mass of the fermion are allowed diverge keeping their ratio, viz., the Fermi velocity finite (i.e. $ k_F, m \rightarrow \infty $ but $ k_F/m = v_F < \infty  $). The RPA limit linearizes the energy momentum dispersion near the Fermi surface ($E=E_F+p v_F$ instead of $E=p^2/(2m)$) \cite{stone1994bosonization}. Units are chosen such that $ \hbar = 1 $ and $ k_F $ is both the Fermi momentum as well as a wavenumber .  For more than one delta potential or a finite barrier/well, etc. it is also essential to define how the width of the impurity cluster `$2 a$' scales in the RPA limit and the assertion is that  $ 2 a k_F   < \infty $ as $ k_F \rightarrow \infty $.  On the other hand, the heights and depths of the various barriers/wells are assumed to be in fixed ratios with the Fermi energy $ E_F = \frac{1}{2} m v_F^2 $ even as $ m \rightarrow \infty $ with $ v_F < \infty $. 

The central quantity that will be used in the calculation of the tunneling conductance is the transmission coefficient ($\tau_0$) of the non-interacting system plus the cluster of impurities which is easily calculated using elementary quantum mechanics and are provided in an earlier work \cite{das2018quantum}. For instance, in the case of a single delta potential: $V_0\delta(x)$,
\begin{equation}
\begin{aligned}
\tau_0=\frac{v_F^2}{V_0^2+v_F^2}
\end{aligned}
\end{equation}
\normalsize
In the case of a double delta potential separated by a distance 2$a$ between them : $V_0( \delta(x+a)+\delta(x-a))$,
\scriptsize
\begin{equation}
\begin{aligned}
\tau_0=\frac{v_F^4}{2V_0^4+2V_0^2v_F^2+v_F^4+2V_0^2(v_F^2-V_0^2)\cos[4k_F a]+4V_0^3v_F\sin[4k_F a]}
\end{aligned}
\end{equation}
\normalsize
For tunneling across a finite barrier: $ V \theta(x+a)\theta(a-x)$ and setting $\lambda = V/E_F$,
\begin{equation}
\begin{aligned}
\tau_0=\frac{8(1-\lambda)}{8-\lambda(8-\lambda)-\lambda^2\cosh[4 k_F a \sqrt{\lambda-1}]}
\end{aligned}
\end{equation} 
When interactions are considered, the (effective) transmission coefficient is modified. Properly defining and expressing this modified tunneling conductance $\tau$ in terms of the non interacting tunneling coefficient $\tau_0$ is the objective of this work.

%------------------------NON CHIRAL BOSONIZATION TECHNIQUE--------------------------------------
\section{ Non chiral bosonization and two point functions}
Analogous to conventional bosonization schemes using the field theoretical approach \cite{giamarchi2004quantum}, the fermionic field operator in NCBT is expressed in terms of currents and densities. But in NCBT, the field operator is modified to include the effect of back-scattering by impurities making it suitable to study translationally non-invariant systems such as the ones considered in this work. The modified field operator of NCBT may be written as follows \cite{das2018quantum}.
\begin{equation}
\begin{aligned}
\psi_{\nu}(x,\sigma,t) \sim C_{\lambda  ,\nu,\gamma}\mbox{ }e^{ i \theta_{\nu}(x,\sigma,t) + 2 \pi i \lambda \nu  \int^{x}_{sgn(x)\infty}\mbox{ } \rho_s(-y,\sigma,t) dy}
\label{PSINU}
\end{aligned}
\end{equation}
Here $\theta_{\nu}$ is the familiar local phase which is a function of the currents and densities. It is also present in the conventional bosonization schemes \cite{giamarchi2004quantum} which goes under the name `g-ology'.
\small
\begin{equation}
\begin{aligned}
\theta_{\nu}(x,\sigma,t) =& \pi \int^{x}_{sgn(x)\infty} dy \bigg( \nu  \mbox{  } \rho_s(y,\sigma,t)\\
&\hspace{1 cm} -  \int^{y}_{sgn(y)\infty} dy^{'} \mbox{ }\partial_{v_F t }  \mbox{ }\rho_s(y^{'},\sigma,t) \bigg)
\end{aligned}
\end{equation}\normalsize
NCBT differs from this by the addition of the optional term $\rho_s(-y)$ in equation (\ref{PSINU}) that ensures the necessary trivial exponents for the single particle Green functions for a system of otherwise free fermions with impurities, which are obtained using standard Fermi algebra. The adjustable parameter $\lambda$, which can take values either 0 or 1, decides the presence or absence of the new term. In other words, setting $\lambda=0$ reduces the NCBT operator  to standard bosonization operator used in g-ology methods. The factor $2 \pi i$ ensures that the necessary fermion commutation rules are obeyed since this term does not change the statistics of the field operator. The quantity $\nu$ signifies a right mover or a left mover and takes values 1 and -1 respectively.
$C_{\lambda  ,\nu,\gamma}$ are pre-factors which can be fixed by comparison using the non-interacting Green functions obtained from Fermi algebra.  The field operator as given in equation (\ref{PSINU}) is to be treated as a mnemonic to obtain the Green functions rather than an operator identity, which avoids the necessity of the Klein factors that are conventionally used. The field operator (annihilation) is clubbed together with another such field operator (creation) and after fixing the C's and $\lambda$'s, one obtains the non interacting two-point functions. Finally the densities $\rho$'s in the RHS of equation (\ref{PSINU}) are replaced by their interacting versions to obtain the many body Green functions, given in \hyperref[AppendixA]{Appendix A}. The details are described in an earlier work \cite{das2018quantum}.

%------------CONDUCTANCE------------------------------------------------------------------------------------------
\section{Conductance }

Conductance may  be thought of as the outcome of a tunneling experiment \cite{kane1992transport}.
Here fermions are injected from one end and collected from the other end. In this sense the conductance is proportional to the magnitude of the effective (i.e. with mutual interactions, possibly at finite temperature) transmission coefficient and is related to the two-point function or the single particle Green function as follows.
\begin{equation}
G = \frac{ e^2 }{h } |T_0| \mbox{   }
| v_F\int^{\infty}_{-\infty}dt\mbox{   }<\{ \psi_{ R } ( \frac{L}{2},\sigma,t) ,  \psi^{\dagger}_{ R } (-\frac{L}{2},\sigma,0)  \}>
 |
 \label{TUNNEL1}
\end{equation}
Here $ |T_0|^2 = \tau_0 $ is the magnitude of the bare transmission coefficient for free fermions plus impurity  in which case the above formula will reduce to the Landauer's formula \cite{das2018transport}. In the presence of interactions,   the results depend on the length of the wire $ L $ and a cutoff $ L_{\omega} = \frac{ v_F }{ k_B T } $ that may be regarded either as inverse temperature or inverse frequency (in case of a.c. conductance). Using the Green function given in \hyperref[AppendixA]{Appendix A}, the conductance can be expressed as a power law as follows (see \cite{das2018transport} for details).
\begin{equation}
G \sim \left( \frac{ L }{ L_{ \omega } }\right)^{\eta } 
\label{GGEN}
\end{equation}
Here $\eta=4X-2Q$. Q and X are given in equation (\ref{luttingerexponents}). In terms of the Luttinger parameter ($g=v_F/v_h$) and the bare transmission coefficient ($\tau_0 = 1-|R_0|^2 = |T_0|^2 $), the exponent can also be written as follows.
\begin{equation}
\eta = \frac{(g-1)(\tau_0 - g - 3 g^2 (1-\tau_0))}{4g(\tau_0 + g(1-\tau_0))}
\label{eta}
\end{equation} 
The conductance given in equation (\ref{GGEN}), obtained using the Green functions in \hyperref[AppendixA]{Appendix A}, is for systems with an infinite bandwidth (more precisely, for temperatures small compared to the bandwidth). The finite bandwidth conductance is discussed in the next section. In this case, the temperature dependence of the tunneling d.c. conductance of a wire with no leads and in the presence of barriers/wells and mutual interaction between particles (forward scattering, infinite bandwidth ie. $ k_F \gg \Lambda_b \rightarrow \infty $) is therefore a simple power-law,
\begin{equation}
G \sim (k_B T)^{ \eta} 
\label{Cond}
\end{equation}
This formula for d.c. conductance is consistent with the assertions of Kane and Fisher \cite{kane1992transport} that show that at low temperatures $ k_B T \rightarrow 0 $ for a fixed $ L $, the conductance vanishes as a power law in the temperature if the interaction between the fermions is repulsive ($ g<1 $). It is also consistent with the infinite bandwidth conductance of weakly interacting electrons in presence of a scatterer of arbitrary strength, as obtained by Matveev et al. \cite{matveev1993tunneling}, the detailed comparison of both the cases been shown in an earlier work \cite{das2018transport}.

%-----------------FINITE BANDWIDTH CONDUCTANCE--------------------------------------------
\section{Finite bandwidth conductance}

The proper way of studying the finite bandwidth conductance would be to re-derive the single particle Green function for finite bandwidth. Also it is important to introduce a bias and calculate the current flowing as a function the bias, temperature, bandwidth etc. and extract the conductance as a linear response coefficient. This has proved to be formidable.  However an acceptable short-cut suggested by referees of our other works is to take the point of view that the transmission and reflection coefficients that appear in $ \eta $ are not the non-interacting temperature independent values but the interacting temperature-dependent values. This amounts to asserting that the equation (\ref{Cond}) which is strictly speaking valid only for temperatures small compared to the bandwidth is now valid in general since $ \eta $ has now been reinterpreted as being temperature and interaction dependent.

Therefore, for electrons with a finite bandwidth $D_0$, the tunneling conductance $\tau$ is given by a transcendental equation viz.
\begin{equation}
\tau = \tau_0 \left ( \frac{k_B T}{D_0} \right)^{\eta(\tau)}
\label{finiteconductance}
\end{equation}
where $\tau_0$ is the tunneling conductance in absence of interactions and the exponents $\eta$ is a function of the conductance $\tau$ and is obtained by replacing the $\tau_0$ in equation (\ref{eta}) by $\tau$.
\begin{equation}
\eta (\tau) = \frac{(g-1)(\tau - g - 3 g^2 (1-\tau))}{4g(\tau + g(1-\tau))}
\label{finiteeta}
\end{equation}  
As before, $ g $ is the Luttinger parameter given by $\frac{v_F}{v_h}$ which is greater than unity for attractive interactions and less than unity for repulsive interactions. $\tau_0$, being the transmission coefficient of the non-interacting system, can be obtained from elementary quantum mechanics and its value ranges from 0 to 1. An exact analytical solution of equation (\ref{finiteconductance}) can't be obtained due to its transcendental nature. However numerical solutions and approximate analytical solutions are possible which are described in the subsequent sub-sections.
%----------------NUMERICAL SOLUTION-------------------------------------------------------------
\subsection{Numerical solution}
The equation (\ref{finiteconductance}) for tunneling conductance may be solved numerically using appropriate empirical values of the remaining parameters.  Based on the transmission coefficient $\tau_0$ and Luttinger liquid parameter $ g $, there are four cases as follows.\\

%-------------------------------------1. weakweak---------------------------------------------------

\begin{figure}[h!]
  \centering
  \includegraphics[scale=0.6]{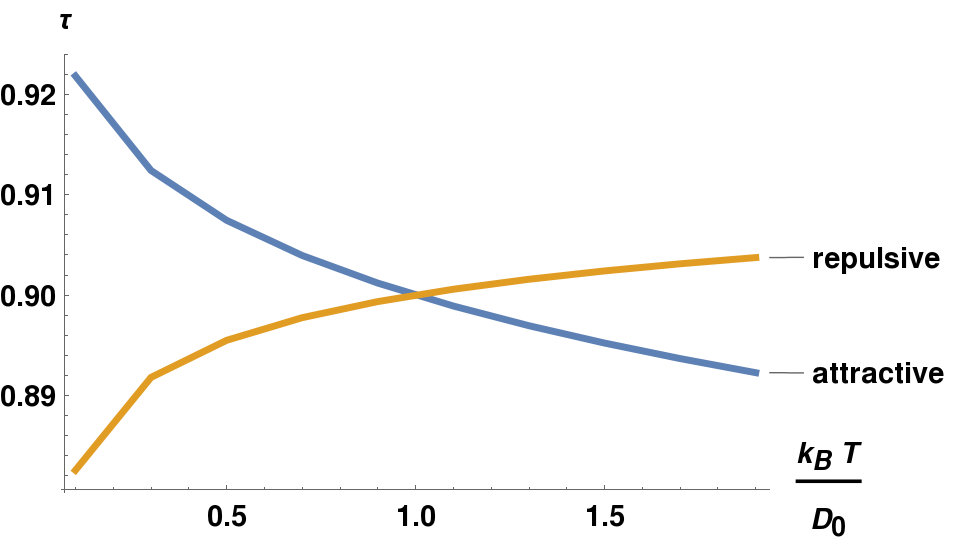}
  \caption{Conductance as a function of  dimensionless temperature ($\frac{k_B T}{D_0}$) for a weak barrier ($\tau_0=0.9$) and weak interactions ($g=0.9$  for repulsive and $g=1.1$ for attractive). }\label{weakweak}
\end{figure}

\noindent{\bf (a) Weak barrier and weak interactions:}
For a weak barrier, there is maximum transmission and hence $\tau_0$ is close to unity. Also for weak interactions, the holon velocity $v_h$ is close to Fermi velocity $v_F$ and hence the following empirical values are chosen:
$\tau_0 = 0.9$; $g = 1.1$ for attractive and $g = 0.9$ for repulsive interactions. For $\frac{k_B T}{D_0}$ ranging from 0.1 to 2, equation (\ref{finiteconductance}) is numerically solved and the obtained values of conductance $\tau$ is plotted as a function of temperature ($\frac{k_B T}{D_0}$) and the graph in figure \ref{weakweak} is obtained.

From the figure \ref{weakweak} it can be seen that near zero temperature, the conductance is close to unity for attractive interactions while it tends to vanish for repulsive interactions. This is the signature of `cutting the chain' by even a small scatterer in case of repulsively interacting particles \cite{kane1992transport}. As the temperature increases, the conductance decreases from its maximum value for attractive interactions while it increases from its minimum value for repulsive interactions. One more observation is that for $k_B T < D_0$, the conductance is larger in the case of attractive interactions while for $k_B T > D_0$, the conductance is larger in the case of repulsive interactions, the transition taking place at the point when $k_B T = D_0$.\\

\begin{figure}[h!]
  \centering
  \includegraphics[scale=0.6]{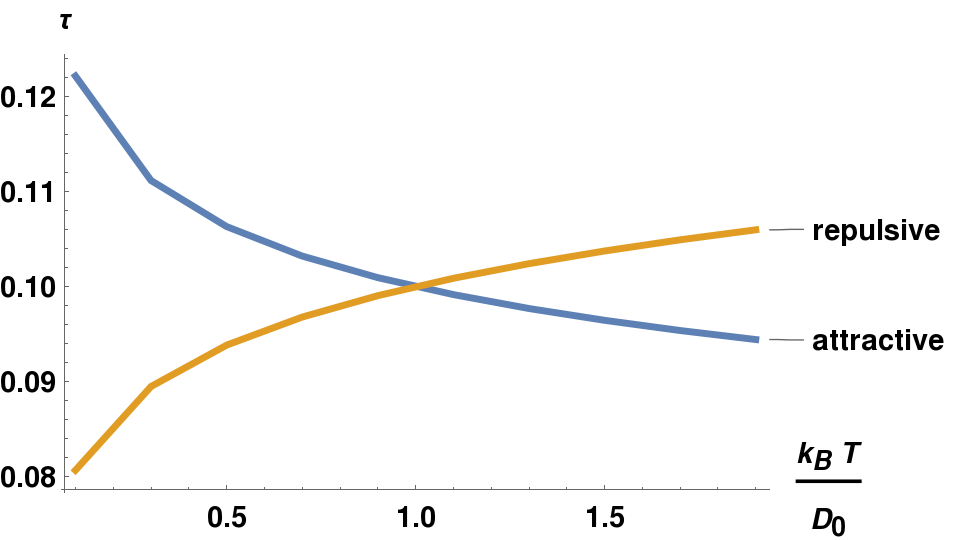}
  \caption{Conductance as a function of  dimensionless temperature ($\frac{k_B T}{D_0}$) for a strong barrier ($\tau_0=0.1$) and weak interactions ($g =$ 0.9 for repulsive and $g =$ 1.1 for attractive). }\label{strongweak}
\end{figure}

%-------------------------------------2. strongweak---------------------------------------------------
\noindent{\bf (b) Strong barrier and weak interactions:}
For a strong barrier, there is minimum transmission and hence $\tau_0$ is close to zero. The following empirical values are chosen:
$\tau_0 = 0.1$; $g = 1.1$ for attractive and $g = 0.9$ for repulsive interactions as in the earlier case. For $\frac{k_B T}{D_0}$ ranging from 0.1 to 2, equation (\ref{finiteconductance}) is numerically solved and the obtained value of conductance $\tau$ is plotted as a function of temperature ($\frac{k_B T}{D_0}$) and the graph in figure \ref{strongweak} is obtained.

Similar observations are made in figure \ref{strongweak} as in the earlier case, the only difference being that for strong barriers conductance is less than that for weak barriers. \\

\begin{figure}[h!]
  \centering
  \includegraphics[scale=0.6]{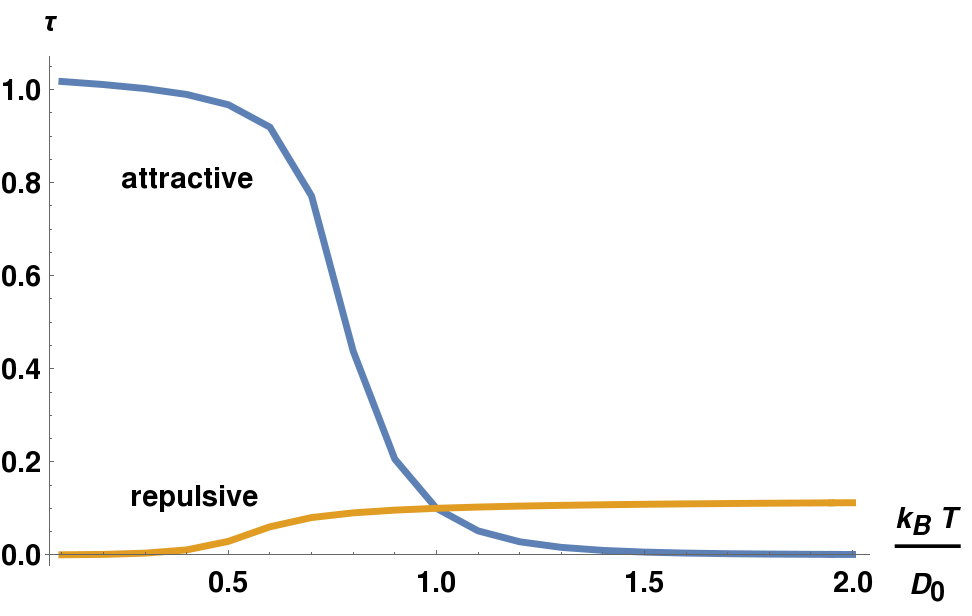}
  \caption{Conductance as a function of  dimensionless temperature ($\frac{k_B T}{D_0}$) for a strong barrier ($\tau_0=0.1$) and strong interactions ($g=$ 0.1 for repulsive and $ g =$ 10 for attractive). }\label{strongstrong}
\end{figure}

%-------------------------------------3. strongstrong---------------------------------------------------
\noindent{\bf (c) Strong barrier and strong interactions:}
When interactions are strong, the holon velocity $v_h$ is quite different from the Fermi velocity $v_F$ and hence the following empirical values are chosen:  $\tau_0 = 0.1$; $g = 10$ for attractive and $g = 0.1$ for repulsive interactions.  Using these values the graph in figure \ref{strongstrong} is obtained.

Figure \ref{strongstrong} clearly signifies the `healing the chain' phenomenon of Kane and Fisher \cite{kane1992transport}. It shows that particles with strong attractive forces between them can tunnel through even the strongest of barriers. This conductance however decreases sharply with an increase in the temperature.

\begin{figure}[h!]
  \centering
  \includegraphics[scale=0.4]{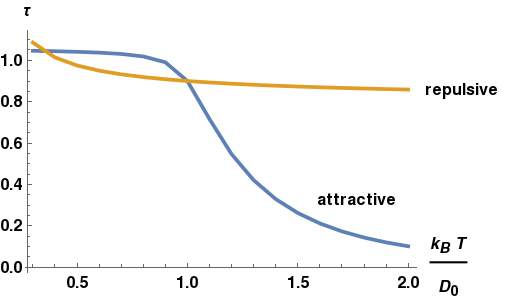}
  \caption{Conductance as a function of  dimensionless temperature ($\frac{k_B T}{D_0}$) for a weak barrier ($\tau_0=0.9$) and strong interactions ($g =$ 0.5 for repulsive and $g =$ 5 for attractive). }\label{weakstrong}
\end{figure}

%-------------------------4. weakstrong---------------------------------------------------
\noindent{\bf (d) Weak barrier and strong interactions:}
The following empirical values are used as a representative of this case:
$\tau_0 = 0.9$; $g = 5$ for attractive and $g = 0.5$ for repulsive interactions.  For $\frac{k_B T}{D_0}$ ranging from 0.3 to 2, equation (\ref{finiteconductance}) is numerically solved and the obtained values of conductance $\tau$ is plotted as a function of temperature ($\frac{k_B T}{D_0}$) and the graph in figure \ref{weakstrong} is obtained.

The plot in this case shows interesting results in the form of high conductance even for repulsive interactions at lower temperatures. 
 It is  well known that for a non homogeneous system, the conductance vanishes at temperatures small compared to bandwidth if the particles are repulsive even if the impurity strength is low, as in the case (a) above. This is believed to be due to a conspiracy between the impurity and mutual interactions which tends to break the chain. But from figure \ref{weakstrong}, it is clear that if the interactions are too strong compared to the strength of the barrier, the system tends to exhibit high conductance at low temperatures rather than exhibiting the well known `cutting the chain' phenomenon of Kane and Fisher \cite{kane1992transport}. However at higher temperatures they are similar to earlier cases. 

\begin{figure}[h!]
  \centering
  \includegraphics[scale=0.6]{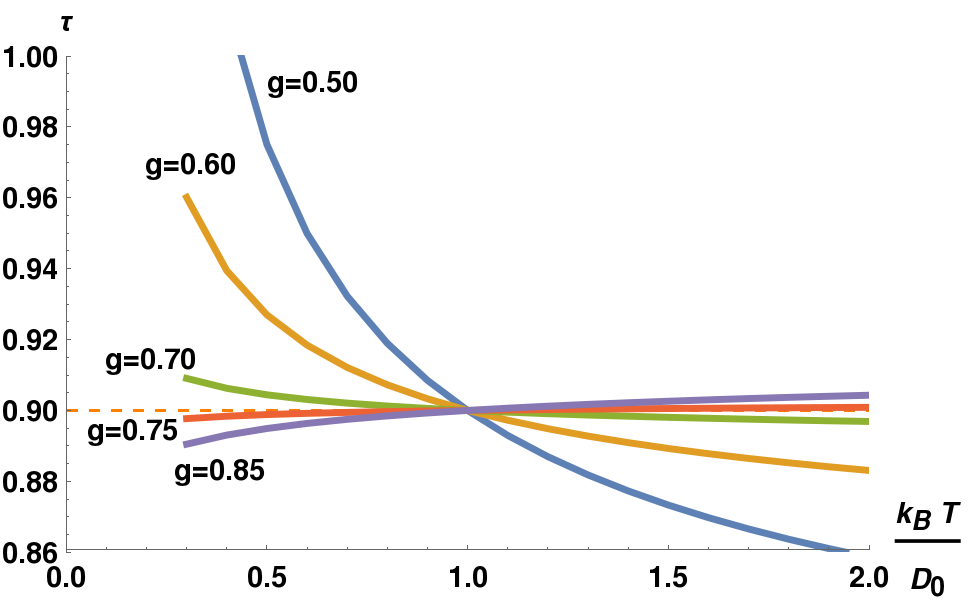}
  \caption{Conductance as a function of  dimensionless temperature ($\frac{k_B T}{D_0}$) for a weak barrier ($\tau_0=0.9$) and various strength of repulsive interactions ($g =$ 0.5 to 0.85). }\label{transition}
\end{figure}

For a weak barrier, the transition from low conductance to high conductance at low temperatures as we increase the strength of repulsions is shown in figure \ref{transition}. This can be understood using an analogy of the conductance of a diode. If one applies a reverse bias to a diode, the conductivity is very less, but if one goes on increasing the reverse bias voltage, at one point it enters into the breakdown region and there is a high flow of current in the reverse direction. Similarly in this case, when there is weak repulsion, the weak barrier behaves like a weak link with low tunneling across it. However when strength of repulsion is increased, it reaches a stage when conductance increases greatly, as with the case with homogeneous LL.
%-------------------MATVEEV COMPARISON-----------------------------------------------------------
\subsection{Comparison with the results of Matveev et al.}
\begin{figure}[h!]
  \centering
  \includegraphics[scale=0.6]{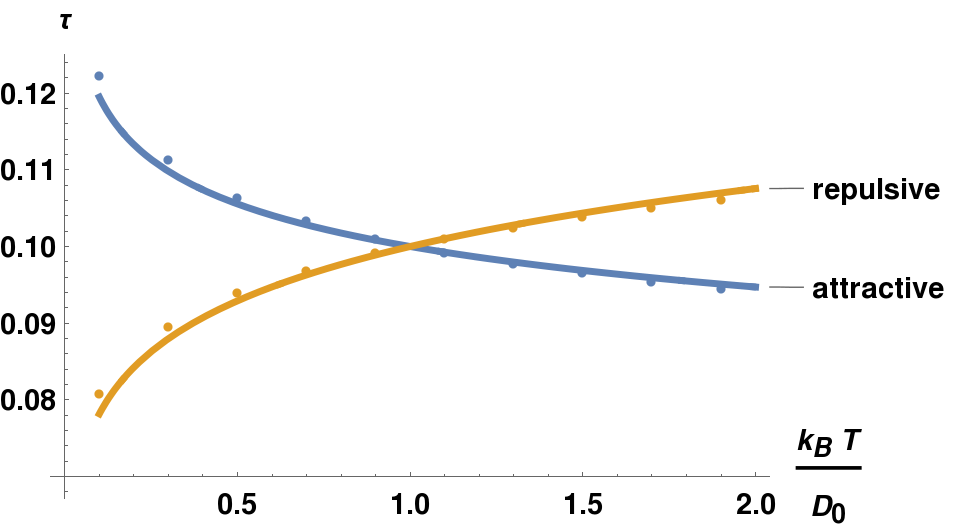}\\
(a)\\
  \includegraphics[scale=0.6]{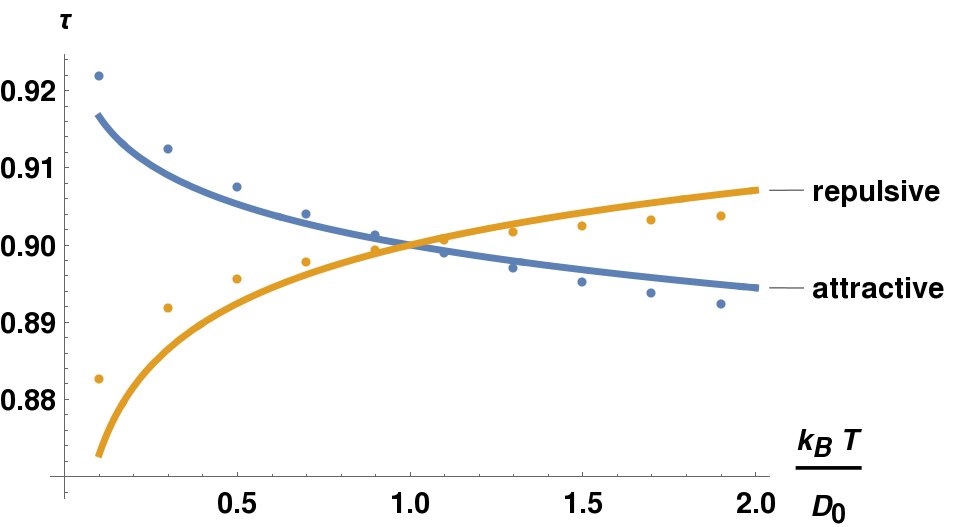}\\
(b)\\
  \caption{Conductance as a function of  dimensionless temperature ($\frac{k_B T}{D_0}$) for weak interactions ($g=$ 0.9 for repulsive and $g=$ 1.1 for attractive).  The dots are numerically exact solution of the transcendental equation obtained from NCBT and the solid line is the analytical formula of Matveev et al. (a) strong  barrier ($\tau_0=0.1$)  (b) weak barrier ($\tau_0=0.9$).}\label{matveevweak}
\end{figure}

The finite bandwidth calculation of conductance as a function of temperature is calculated by Matveev et al. \cite{matveev1993tunneling} and is given in equation (14) of their paper as follows (setting $e^2/h = 1$ to tally with our results).
\begin{equation}
G(T) = \frac{\tau_0 \left(\frac{k_B T}{D_0}\right)^{2\alpha}}{1-\tau_0 + \tau_0 \left(\frac{k_B T}{D_0}\right)^{2\alpha}}
\label{Matveev14}
\end{equation}
and using their terminology for forward scattering interactions only, we have
\begin{equation}
2 \alpha = \frac{v_0}{ \pi v_F}
\label{alphasmall}
\end{equation}
Expressing $\alpha$ in terms of the Luttinger parameter $ g $ used in this work, $2 \alpha = \frac{1}{2}\left(\frac{1}{g^2} - 1\right)$. The formula in the equation (\ref{Matveev14}) is valid only for weak interactions, as claimed by the authors \cite{matveev1993tunneling}. So the comparison is done only for weak interactions and hence the empirical values of Luttinger parameter $g$ is chosen to be 0.9 for repulsive interactions and 1.1 for attractive interactions. The conductance obtained for both strong barrier ($\tau_0=0.1$) and weak barrier ($\tau_0=0.9$) obtained using the analytical formula in equation (\ref{Matveev14}) and that obtained using numerical solution of our results are plotted as a function of temperature ($\frac{k_B T}{D_0}$) in figure \ref{matveevweak}.
%-----------------------fig 5 -----------------------------------

In figure \ref{matveevweak}, the continuous lines are obtained from the analytical formulas of conductance by Matveev et al. while the dots represents the numerical solution of the conductance obtained in the present work. It is seen that they are in good agreement with each other. For temperature very close to zero ($\frac{k_B T}{D_0}=0.01$), the values of conductance obtained by both the methods (Matveev et al. and NCBT) are also in good agreement with each other. The following empirical values are chosen: $\tau_0$ is 0.9 for weak barrier and 0.1 for strong barrier while $ g $ is 1.1 for attractive and 0.9 for repulsive interactions.
The comparison is shown in the table \ref{neartozero}.

%------------
%-----------------------------------------------table 1---------------------------------------------------------------------------
\noindent\begin{table}[h!]
\caption{  Values of  conductance $ \tau $ for weak interactions as obtained by Matveev et al. and NCBT for temperature very close to  zero ($\frac{k_B T}{D_0}=0.01$). }
{\begin{tabular}{c c c }
\hline
Case \hspace{3 cm} & Matveev et al. \hspace{1 cm}& NCBT\hspace{1 cm}\\	[5pt]
\hline
Weak barrier + attraction & 0.930 &0.937 \\
Weak barrier + repulsion & 0.840&0.854 \\
Strong barrier + attraction &0.142 & 0.148\\
Strong barrier + repulsion & 0.061& 0.064\\
\hline
\end{tabular}
}
\label{neartozero}
\end{table}

On the other hand, for temperatures much greater than the bandwidth ($\frac{k_B T}{D_0}=50$), the values of conductance obtained by both the methods (Matveev et al. and NCBT) are again in good agreement with each other. The same empirical values are chosen: $\tau_0$ is 0.9 for weak barrier and 0.1 for strong barrier while $ g $ is 1.1 for attractive and 0.9 for repulsive interactions. The comparison is shown in the table \ref{farfromzero}. As temperature is further increased the strength of interactions has to be decreased for a more favorable comparison.

%-----------------------------------------------table 2---------------------------------------------------------------------------
\noindent\begin{table}[h!]
\caption{  Values of  conductance $ \tau $ for weak interactions as obtained by Matveev et al. and NCBT for temperature much greater than bandwidth ($\frac{k_B T}{D_0}=50$). }
{\begin{tabular}{c c c }
\hline
Case \hspace{3 cm} & Matveev et al. \hspace{1 cm}& NCBT\hspace{1 cm}\\	[5pt]
\hline
Weak barrier + attraction & 0.865 &0.834 \\
Weak barrier + repulsion & 0.934&0.918 \\
Strong barrier + attraction &0.073& 0.070\\
Strong barrier + repulsion & 0.150& 0.140\\
\hline
\end{tabular}
}
\label{farfromzero}
\end{table}

The central equation of finite bandwidth conductance of Matveev et al. valid for weak interactions as given by equation (\ref{Matveev14}) can actually be obtained by considering the weak interaction limit of the transcendental equation (\ref{finiteconductance}) obtained using NCBT which is valid for any strength of interactions. Setting $y=\frac{k_B T}{D_0}$, equation (\ref{finiteconductance}) reads as follows.
\begin{equation*}
\tau = \tau_0 \mbox{ }y^{\eta(\tau)}
\end{equation*}
Differentiating with respect to y,
\begin{equation*}
\frac{d \tau}{d y} = \tau_0 \mbox{ } \eta(\tau)\mbox{ }y^{\eta(\tau)-1} + \tau_0 \mbox{ } \log[y]\mbox{ } y^{\eta(\tau)}\frac{d\eta(\tau)}{dy}
%= \tau_0 \mbox{ }y^{\eta(\tau)} \eta(\tau)\mbox{ }\frac{1}{y}= \eta(\tau) \frac{\tau}{y} 
\end{equation*}
For weak interactions, $\eta(\tau)=\frac{v_0}{\pi v_F}(1-\tau) = 2 \alpha (1-\tau) $,
\begin{equation*}
\begin{aligned}
&\frac{d \tau}{d y} =  2 \alpha (1-\tau) \frac{\tau}{y} - \tau \mbox{ } \log[y] \mbox{ }2 \alpha \frac{d \tau}{d y}\\
\end{aligned}
\end{equation*}
For weak interactions, $2\alpha$ is very small and hence one can write,
\begin{equation*}
\begin{aligned}
\frac{d \tau}{d y} =   \frac{\tau(1-\tau)}{y} \frac{ 2 \alpha }{ (1+2 \alpha\mbox{ }\tau \mbox{ } \log[y] \mbox{ }) } 
 \approx 2 \alpha \mbox{   }  \frac{\tau(1-\tau)}{y}
\end{aligned}
\end{equation*}
%So the following differential equation has to solved.
%\begin{equation*}
%\frac{d \tau}{\tau (1-\tau)} =  2 \alpha  \frac{dy}{y} 
%\end{equation*}
Using appropriate limits of integration,
\begin{equation*}
\log\left(\frac{ \tau}{ 1-\tau}\right) \Big|_{\tau_0}^{\tau} =  2 \alpha  \log(y) \Big|_{1}^{y}
\end{equation*}
which gives,
\begin{equation*}
\frac{\tau(1-\tau_0)}{\tau_0(1-\tau)} = y^{2\alpha}
\end{equation*}
which may be easily solved for $\tau$ (replacing y by $\frac{k_BT}{D_0}$),
\begin{equation}
\tau=\frac{\tau_0 \left(\frac{k_B T}{D_0}\right)^{2\alpha}}{1-\tau_0 + \tau_0 \left(\frac{k_B T}{D_0}\right)^{2\alpha}}
\end{equation}
which is precisely equation (\ref{Matveev14}) obtained by Matveev et al. \cite{matveev1993tunneling} in their work.

%\[
%1+(\frac{v_0}{\pi v_F}) \tau \log{[k_B T/D_0]}
%\]

\begin{figure}[h!]
  \centering
  \includegraphics[scale=0.45]{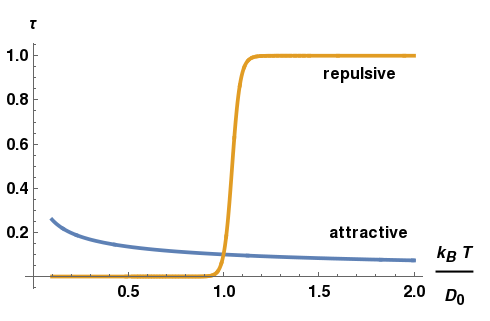}
  \caption{Conductance as a function of  dimensionless temperature ($\frac{k_B T}{D_0}$) for a strong barrier ($\tau_0=0.1$) and strong interactions ($g=$ 0.1 for repulsive and $ g =$ 10 for attractive) using Matveev et al.'s formula. }\label{matveevwrong}
\end{figure}

As claimed by Matveev et al. \cite{matveev1993tunneling}, the formula as given in equation (\ref{Matveev14}) is valid for weak interactions. The breakdown of the Matveev et al.'s formula for conductance at strong interactions can be seen from figure \ref{matveevwrong}. Choosing empirical values for a strong barrier ($\tau_0 = 0.1$) and strong interactions ($g=0.1$ for repulsive and $g=10$ for attractive), the conductance is plotted as a function of temperature and the graphs in figure \ref{matveevwrong} are obtained. In the case (c) of the previous subsection, it has been shown how NCBT conductance, for the exact same case as above, supports the healing the chain phenomenon for attractive interactions. But the figure \ref{matveevwrong} somewhat violates the cutting the chain phenomenon for strong repulsive interactions, that too with a strong barrier.
%-------------------ANOMALOUS CONDUCTANCE-------------------------------------------------------
\subsection{Anomalous conductance}

\begin{figure}[h!]
  \centering
  \includegraphics[scale=0.6]{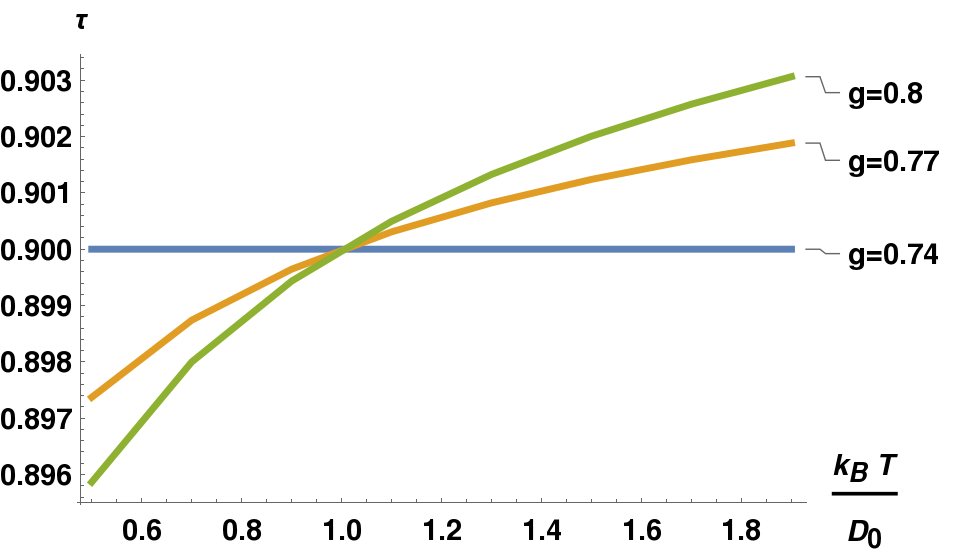}\\
(a)\\
  \includegraphics[scale=0.6]{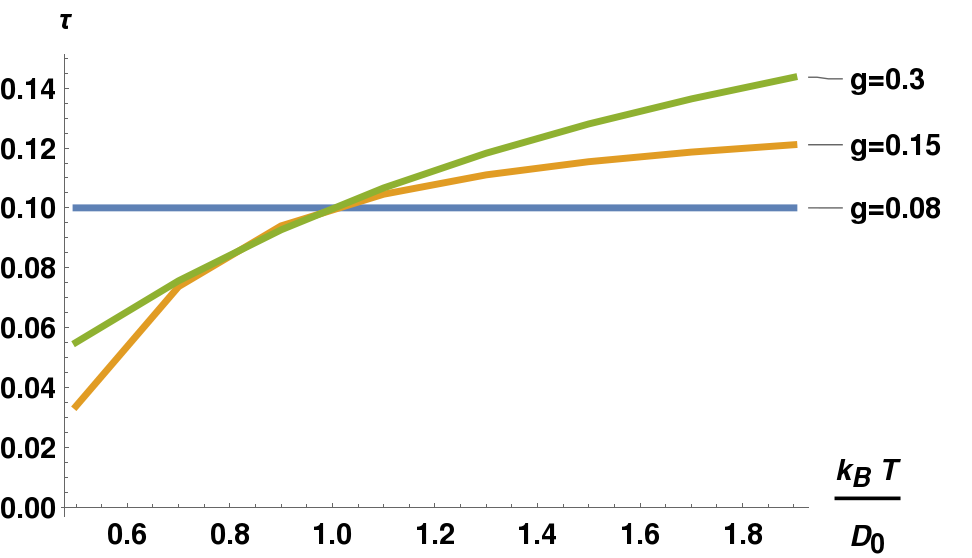}\\
(b)\\
  \caption{Conductance as a function of  dimensionless temperature ($\frac{k_B T}{D_0}$) for values of $g$ near $g_0$ where conductance exponent $\eta$ vanishes: (a) weak barrier ($\tau_0=0.9$) with $g_0=0.74$. (a) strong barrier ($\tau_0=0.9$) with $g_0=0.08$}\label{anoweak}
\end{figure}

It has been observed from the earlier plots that with an increase in temperature, conductance typically decreases for attractive interactions and increases for repulsive interactions. This is because the exponent $\eta$ in equation (\ref{finiteconductance}) is typically positive for repulsive cases and negative for attractive cases. But in absence of interactions ($g=$1), the exponent $\eta$ vanishes and the conductance becomes independent of temperature and is given by 
\[
\tau_{\eta=0} = \tau_0 
\]
However $g=1$ is not the only condition for which $\eta$ vanishes, the other condition being
\[
g_0= \frac{-1 \pm \sqrt{1+12 \tau_0 - 12 \tau_0^2}}{6(1-\tau_0)}
\] 
Since $0\le \tau_0 \le 1$, hence $\sqrt{1+12 \tau_0 - 12 \tau_0^2} \ge 1$. But since $g $ can't be negative ($g = v_F/v_h$), the only admissible value of $g$ for $\eta=0$ is
\[
g_0= \frac{-1 + \sqrt{1+12 \tau_0 - 12 \tau_0^2}}{6(1-\tau_0)}
\]
Here $g_0$ is the value of $g$ for which $\eta$ vanishes.
In presence of  a strong barrier ($\tau_0 = 0.1$) the conductance becomes temperature independent ($\eta = 0$) for $g_0=0.08$ which indicates very strong repulsion. On the other hand for a weak barrier ($\tau_0 = 0.9$) this happens for $g_0=0.74$ which is also repulsive but less stronger. This can be thought of as a conspiracy between the impurity and the repulsive interactions to give rise to a state that is similar to the non-interacting one. In figure \ref{anoweak}, the conductance is shown as a function of temperature for values of $g$ near  $g_0$. It can be seen that as $g$ approaches $g_0$, the temperature dependence of conductance becomes weaker and weaker (the graph flattens) and finally becomes independent (constant graph) for $g=g_0$.
%-----------------------fig 6 -----------------------------------

%------------------------ANALYTICAL SOLUTION----------------------------------------------------
\subsection{Analytical solution}
\begin{figure}[h!]
  \centering
  \includegraphics[scale=0.6]{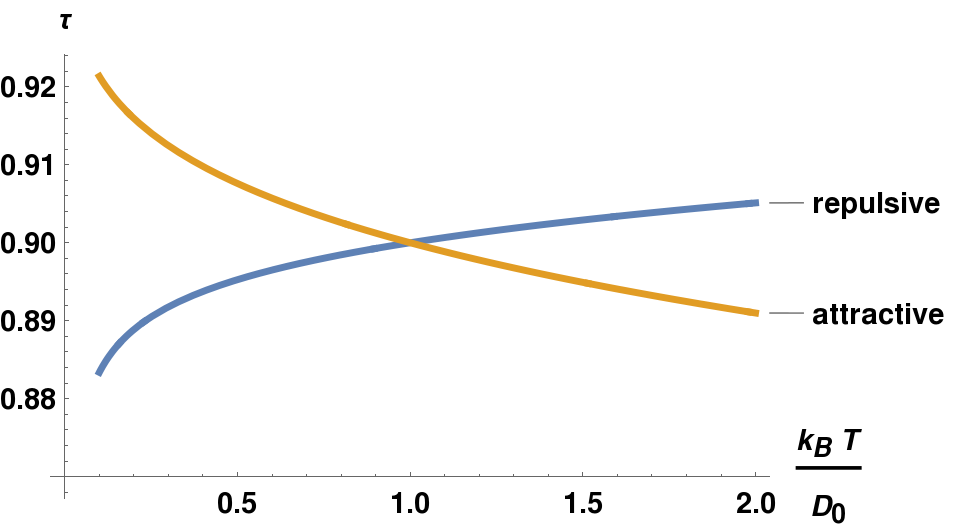}\\
(a)\\
  \includegraphics[scale=0.6]{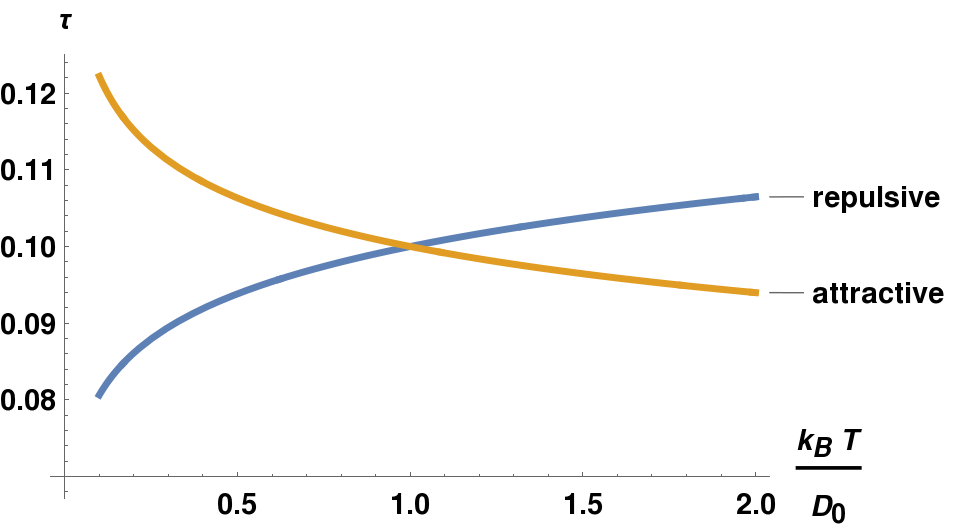}\\
(b)
  \caption{Conductance from analytical expression as a function of  dimensionless temperature ($\frac{k_B T}{D_0}$) for weak interactions (g $=$ 0.9 for repulsive and $g=$ 1.1 for attractive): (a) weak barrier ($\tau_0=0.9$) (b) strong barrier ($\tau_0=0.1$).
  }\label{analyticalweak}
\end{figure}

Equation (\ref{finiteconductance}) being transcendental in nature can't be solved analytically. However using the fact that the tunneling conductance $\tau$ is less than unity, the RHS of the equation can be expanded in powers of $\tau$ and truncated after a certain order. Smaller the value of $\tau$, sooner can the series be truncated. Ignoring the third and higher powers of the series and solving the rest of the equation, the following expression of tunneling conductance is obtained.\footnotesize
\begin{equation}
\tau = \frac{-4 g^2 (\frac{k_B T}{D_0})^{\frac{1}{4}(2+\frac{1}{g})}\tau_0}{(g^2-1)(\frac{k_B T}{D_0})^{\frac{1}{4}(2+\frac{1}{g})}\tau_0 \log(\frac{k_B T}{D_0})-g^2(\frac{k_B T}{D_0})^{\frac{3g}{4}}(2+\sqrt{C})}
\label{anatau}
\end{equation}\normalsize
where \footnotesize
\begin{equation*}
\begin{aligned}
C=&4-4(\frac{k_B T}{D_0})^{\frac{1}{4}(2+\frac{1}{g}-3g)}\tau_0 \log(\frac{k_B T}{D_0})(1-\frac{1}{g^2})\\
-&\frac{1}{g^4}\Big((1-g)^2(1+g)(\frac{k_B T}{D_0})^{\frac{1}{4}(2+\frac{1}{g}-3g)}\tau_0^2 \log(\frac{k_B T}{D_0})\\
&\times(8g+(1+g)\log(\frac{k_B T}{D_0}))\Big)
\end{aligned}
\end{equation*}\normalsize
Using the analytical expression in equation (\ref{anatau}) for weak interactions ($g=$ 0.9, 1.1) the conductance is plotted as a function of temperature ($\frac{k_B T}{D_0}$) for both weak ($\tau_0=0.9$) and  strong ($\tau_0=0.1$) barriers in figures \ref{analyticalweak}(a)  and \ref{analyticalweak}(b) respectively and they are in close agreement with those obtained for the results of Matveev et al. depicted in  figures \ref{matveevweak}(a) and \ref{matveevweak}(b) respectively.

%-----------------------fig 7 -----------------------------------
%-------------------------------------------
The conductance equation of Matveev et al. given by equation (\ref{Matveev14}) may be expanded in powers of the interaction parameter and terms retained up to the first order (since it is for weak interactions). On the other hand, the analytical expression of conductance from NCBT given by equation (\ref{anatau}) can also be expanded in terms of the interaction parameter and terms may be retained up to the first order. In both cases, the following is obtained which is an exact match for finite temperature conductance for weak interactions ($g \sim$ 1).
\begin{equation*}
\tau=\tau_0 + \frac{(1-g^2)\tau_0(1-\tau_0)\log(\frac{k_B T}{D_0})}{2g^2}
\end{equation*}
%-------------------------COMPARISON ANALYTICAL NUMERICAL------------------------------------------
\subsection{Comparison of analytical and numerical solution}
The analytical solution using the second order approximation to the conductance and the numerical solution of the exact transcendental equation has to be compared so that one can estimate how good the approximation is for various cases. Choosing the empirical values of $g=$ 0.9, 1.1 for weak interactions, $g=$ 0.3, 3 for strong interactions and $\tau_0$ to be 0.1 for strong barrier and 0.9 for weak barriers, the values of conductance are compared for different combinations of $g$'s and $\tau_0$'s for the case  $\frac{k_B T}{D_0} = 0.5$ and the results are tabulated in table \ref{numanacomp}.

%-----------------------------------------------table 1---------------------------------------------------------------------------
\noindent\begin{table}[h!]
\caption{  Comparison of the values of  conductance $ \tau $ obtained numerically and analytically for $\frac{k_B T}{D_0} = 0.5$ }
{\begin{tabular}{c c c }
\hline
Case \hspace{3 cm} & Numerical \hspace{1 cm}& Analytical\hspace{1 cm}\\	[5pt]
\hline
Strong barrier + strong attraction  &0.285 & 0.285\\
Strong barrier + weak attraction &0.106 & 0.106\\
Strong barrier + strong repulsion  &0.055 & 0.055\\
Strong barrier + weak repulsion  &0.094 & 0.094\\
Weak barrier + strong attraction &1.032 &1.183 \\
Weak barrier + weak attraction &0.907 &0.908 \\
Weak barrier + strong repulsion &1.288 & 1.024\\
Weak barrier + weak repulsion &0.895 & 0.895\\
\hline
\end{tabular}
}
\label{numanacomp}
\end{table}

From table \ref{numanacomp} it can be observed that for strong barrier ($\tau_0 \sim 0$), the numerical and analytical values are precisely matching as ignoring the higher powers of $\tau_0$ is a much better approximation in this case. For weak barriers ($\tau_0 \sim 1$) ignoring the higher powers of $\tau_0$ is less accurate, especially for attractive interactions which tends to mitigate the effect of the barrier (healing the chain phenomenon as described by Kane and Fisher \cite{kane1992transport}) and make $\tau_0$ approach unity. Thus there is a minor mismatch between the analytical and the numerical values for weak attractions and a little more for strong attractions. On the other hand repulsive interactions aggravate the effect of the barrier (cutting the chain phenomenon of Kane and Fisher \cite{kane1992transport}) and minimize the tunneling. Thus for weak repulsion, there is an exact match between the two values inspite of weak barriers. However for strong repulsions and weak barrier, there is some anomalous behavior as also depicted in figure \ref{weakstrong} above, where the value of conductance tends towards unity and hence the second order approximation is not a very good one for this case.

%-------------------------BACKWARD SCATTERING-----------------------------------------

\subsection{Both forward and backward scattering}
The transcendental equation given by the equation (\ref{finiteconductance}) and expression of the exponent $\eta(\tau)$ given by equation (\ref{finiteeta}) remains the same upon inclusion of backward scattering interactions between fermions. The difference comes in the expression of the holon velocity $v_h$ (note that $g=v_F/v_h$), which is now modified to include the effect of backward scattering. Considering $v_0$ is the strength of forward scattering interactions as discussed in equation (\ref{vq}), the holon velocity is given by $v_h =  v_F \sqrt{1+ \frac{2 v_0}{\pi v_F}}$. In presence of backward scattering (of strength $v_1$) the $v_0$ is replaced by an effective $v_0$ given by (in this work we only deal with fermions with spin) 
\begin{equation}
v_{0,eff} =  g_2(T) - 2 g_1 (T)
\label{v0eff}
\end{equation}
where $g_1$ and $g_2$ are the renormalized values of backward and forward scattering interaction strengths that can be derived using Parquet's approximation \cite{solyom1979fermi} and are given by
\begin{equation}
\begin{aligned}
g_1(T) = &\frac{v_1}{1+\frac{v_1}{\pi v_F}\log{[\frac{D_0}{k_B T}]}}\\
g_2(T) = &v_0 - \frac{v_1}{2} + \frac{v_1}{2(1+\frac{v_1}{\pi v_F}\log{[\frac{D_0}{k_B T}]})}\\
\label{g1g2}
\end{aligned}
\end{equation}
Hence the Luttinger parameter $g$ used in equation (\ref{finiteeta}) can be written for small values of interactions as follows.
\begin{equation}
g=\frac{1}{\sqrt{1+\frac{2 v_{0,eff}}{\pi v_F}}} \approx 1-\frac{ v_{0,eff}}{\pi v_F}
\end{equation}
Setting $y=\frac{k_B T}{D_0}$, equation (\ref{finiteconductance}) reads as follows.
\begin{equation*}
\tau = \tau_0 \mbox{ }y^{\eta(\tau)}
\end{equation*}
Differentiating with respect to y,
\begin{equation*}
\frac{d \tau}{d y} = \tau_0 \mbox{ } \eta(\tau)\mbox{ }y^{\eta(\tau)-1} + \tau_0 \mbox{ } \log[y]\mbox{ } y^{\eta(\tau)}\frac{d\eta(\tau)}{dy}
\end{equation*}
For weak interactions, $\eta(\tau)=\frac{v_{0,eff}}{\pi v_F}(1-\tau)$. Setting $\frac{v_{0,eff}}{\pi v_F}=2\alpha_{eff}$,
\begin{equation*}
\begin{aligned}
&\frac{d \tau}{d y} =  2 \alpha_{eff} (1-\tau) \frac{\tau}{y} - \tau \mbox{ } \log[y] \mbox{ }2 \alpha_{eff} \frac{d \tau}{d y}\\
\end{aligned}
\end{equation*}
For weak interactions, $2\alpha_{eff}$ is very small and hence one can write,
\begin{equation*}
\begin{aligned}
\frac{d \tau}{d y} =   \frac{\tau(1-\tau)}{y} \frac{ 2 \alpha_{eff} }{ (1+2 \alpha_{eff}\mbox{ }\tau \mbox{ } \log[y] \mbox{ }) } 
 \approx 2 \alpha_{eff} \mbox{   }  \frac{\tau(1-\tau)}{y}
\end{aligned}
\end{equation*}
Using equations (\ref{v0eff}) and (\ref{g1g2}), 
\[
\alpha_{eff} = \alpha_2 + \alpha_1 \left(-\frac{1}{2} - \frac{3}{2 - 4 \alpha_1 \log{[y]}}\right)
\]
Hence the differential equation takes the form
\[
\frac{d \tau}{d y} =  2 \left( \alpha_2 + \alpha_1 \left(-\frac{1}{2} - \frac{3}{2 - 4 \alpha_1 \log{[y]}}\right)\right) \mbox{   }  \frac{\tau(1-\tau)}{y}
\]
which is solved and using appropriate limits of integration,
\begin{equation*}
\tau = \frac{\tau_0 \left( 1 + 2 \alpha_1 \log{(\frac{D_0}{k_BT})} \right)^{\frac{3}{2}}\left(\frac{k_BT}{D_0}\right)^{2\alpha_2-\alpha_1} }{1-\tau_0+\tau_0 \left( 1 + 2 \alpha_1 \log{(\frac{D_0}{k_BT})} \right)^{\frac{3}{2}}\left(\frac{k_BT}{D_0}\right)^{2\alpha_2-\alpha_1}}
\end{equation*}
which is the conductance for weakly interacting electrons with both backward and forward scattering as also given by equation (21) of Matveev et al. \cite{matveev1993tunneling}.

\begin{figure}[h!]
  \centering
  \includegraphics[scale=0.4]{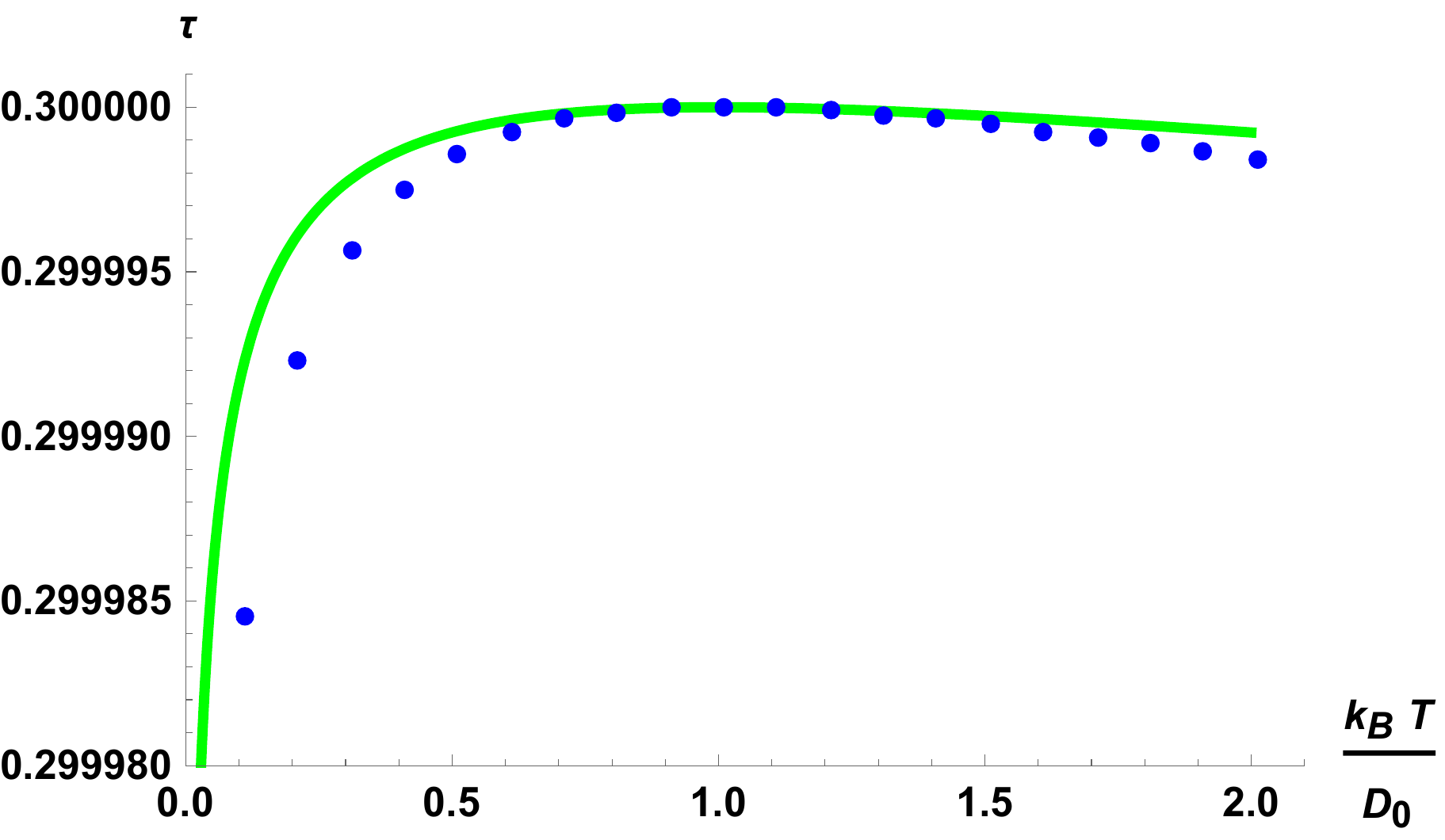}
  \caption{Conductance as a function of  dimensionless temperature ($\frac{k_B T}{D_0}$) for weak interactions ( $v_F=1$, $v_0 = 0.02$ and $v_1 = 0.01$) and a barrier of strength $\tau_0 = 0.3$. The dots represent the NCBT numerical solution and the continuous line represents the analytical solution of Matveev et al.}
\label{backwardsmall}
\end{figure}

It is interesting to see the interplay between the forward and backward scattering interactions. For weak or no backward scattering, the conductance shows a monotonic behavior with respect to temperature. When backward scattering is increased gradually, the conductance starts showing a non-monotonic behavior such that with an increase in temperature, the conductance first increases, reaches a maximum and then decreases. This is shown in  figure \ref{backwardsmall} by numerically solving the transcendental equation (\ref{finiteconductance}) for the empirical values of $\tau_0 = 0.3$, $v_F = 1$, $v_0 = 0.02$ and $v_1 = 0.01$. The solution (depicted by the dots) is in good agreement with that of Matveev et al. (continuous line).

%-------------------------Monte carlo---------------------------------------

\subsection{Comparison with Monte Carlo results}
The conductance of Luttinger liquids with impurities has been studied using numerical methods like Monte Carlo simulations \cite{hamamoto2008numerical, morath2016conductance, leung1995dynamical}. In the work by Hamamoto et al. \cite{hamamoto2008numerical}, where path integral Monte Carlo methods are used, it has been found that the d.c. conductance increases monotonically for $K_{\rho}>1.025$ ($K_{\rho} = g$ in our notation and hence weakly attractive) with a decrease in temperature. Whereas for $K_{\rho}<0.975$ (weakly repulsive) it decreases monotonically with a decrease in temperature. This is in good agreement with the plots in figures \ref{weakweak} and \ref{strongweak} which are also for weak interactions ($g=1.1$ for attractive and $g=0.9$ for repulsive). Similar trends were earlier obtained by Leung et al. \cite{leung1995dynamical} for repulsive interactions with $g=1/3$ and $g=1/6$. 

%----DISCUSSION-------------------------------------------------------------
\section{Discussion}

The approach of the present work is arrived at as follows. First we make the following observations. 

i) The work of Matveev, Yue and Glazman \cite{matveev1993tunneling} is valid for both forward and backward scattering between fermions and also for finite bandwidth. However, it is only valid for weak coupling between fermions whether it is attractive or repulsive.

ii) The NCBT of our original work \cite{das2018quantum} is valid for arbitrary strengths of interactions. However, it is only applicable for large (compared to temperature) bandwidth and only for forward scattering between fermions.

The goal of the present work is to find an overarching formalism that subsumes both cases i) and ii) and reproduces them as appropriate limiting cases. Also the idea is to keep the formalism simple and believable. It is the claim of the present work that this goal has been achieved.

%----CONCLUSION-------------------------------------------------------------
\section{Conclusions}
In this work, the correlation functions of a Luttinger liquid with impurities obtained using the newly constructed non-chiral bosonization technique (NCBT) are used to calculate the tunneling conductance as a function of temperature for forward-scattering  mutual interaction between fermions with a finite bandwidth. The results are valid for arbitrary strength of the impurities as well as that of interactions and compare favorably with those obtained by Matveev et al. for weakly interacting systems. Novel physics in the form of a weakly temperature dependent conductance is seen when the mutual repulsion between fermions is large and the holon velocity bears a certain well-defined relation to the bare transmission coefficient of the system. Deviations from the `cutting the chain' phenomenon is observed for a weak scatterer when the strength of repulsion is strong, leading to an unusual high conductance at lower temperatures, similar to breakdown current in a diode. Upon inclusion of backward scattering, there occurs an interplay between the forward and backward scattering such that the monotonic temperature dependence which dominates for forward scattering starts showing non-monotonic behavior when backward scattering is gradually increased.

%-------------------APPENDIX I-----------------------------------------------------------
\section*{APPENDIX A:  Two point functions using NCBT}
\label{AppendixA}
\setcounter{equation}{0}
\renewcommand{\theequation}{A.\arabic{equation}}
The full single particle Green function of a Luttinger liquid in presence of impurities has been derived using the NCBT in an earlier work \cite{das2018quantum}. This Green function has been shown to obey both conventional perturbation theory as well as the exact Scwinger Dyson equation \cite{das2018non}.

The full Green function is the sum of all the parts. The notion of weak equality is introduced which is denoted by \begin{small} $ A[X_1,X_2] \sim B[X_1,X_2] $ \end{small}. This really means  \begin{small} $ \partial_{t_1} Log[ A[X_1,X_2] ]  = \partial_{t_1} Log[ B[X_1,X_2] ] $\end{small} assuming that A and B do not vanish identically. {\bf Notation:} $X_i \equiv (x_i,\sigma_i,t_i)$ and  $\tau_{12} =  t_1 - t_2$. 
\scriptsize

\begin{equation}
\begin{aligned}
\Big\langle T\mbox{  }\psi(X_1)\psi^{\dagger}(X_2) \Big\rangle 
=&\Big\langle T\mbox{  }\psi_{R}(X_1)\psi_{R}^{\dagger}(X_2) \Big\rangle +\Big \langle T\mbox{  }\psi_{L}(X_1)\psi_{L}^{\dagger}(X_2)\Big\rangle \\
+&\Big\langle T\mbox{  }\psi_{R}(X_1)\psi_{L}^{\dagger}(X_2) \Big\rangle + \Big\langle T\mbox{  }\psi_{L}(X_1)\psi_{R}^{\dagger}(X_2)\Big\rangle \\
\label{break}
\end{aligned}
\end{equation}

\small
\begin{bf} Case I : $x_1$ and $x_2$ on the same side of the origin\end{bf} \\ \scriptsize

\begin{equation*}
\begin{aligned}
\Big\langle T\mbox{  }\psi&_{R}(X_1)\psi_{R}^{\dagger}(X_2)\Big\rangle \sim 
\frac{(4x_1x_2)^{\gamma_1}}{(x_1-x_2 -v_h \tau_{12})^{P} (-x_1+x_2 -v_h \tau_{12})^{Q}} \\
\times&\frac{1}{ (x_1+x_2 -v_h \tau_{12})^{X} (-x_1-x_2 -v_h \tau_{12})^{X} (x_1-x_2 -v_F \tau_{12})^{0.5}}\\
%%%%%%%%%%%%%%%%%%%%%
\Big\langle T\mbox{  }\psi&_{L}(X_1)\psi_{L}^{\dagger}(X_2)\Big\rangle \sim 
\frac{(4x_1x_2)^{\gamma_1}}{(x_1-x_2 -v_h \tau_{12})^{Q} (-x_1+x_2 -v_h \tau_{12})^{P}} \\
\times&\frac{1}{ (x_1+x_2 -v_h \tau_{12})^{X} (-x_1-x_2 -v_h \tau_{12})^{X}(-x_1+x_2 -v_F \tau_{12})^{0.5}}\\
%%%%%%%%%%%%%%%%%%%%%
\end{aligned}
\end{equation*}

%%%%%%%%%%%%%%%%%%%%%%%%%%%%%%%%%%%%%%%%%%%%%%%
\begin{equation}
\begin{aligned}
\Big\langle T\mbox{  }\psi&_{R}(X_1)\psi_{L}^{\dagger}(X_2)\Big\rangle \sim 
\frac{(2x_1)^{\gamma_1}(2x_2)^{1+\gamma_2}+(2x_1)^{1+\gamma_2}(2x_2)^{\gamma_1}}{2(x_1-x_2 -v_h \tau_{12})^{S} (-x_1+x_2 -v_h \tau_{12})^{S}} \\
\times&\frac{1}{ (x_1+x_2 -v_h \tau_{12})^{Y} (-x_1-x_2 -v_h \tau_{12})^{Z}(x_1+x_2 -v_F \tau_{12})^{0.5}}\\
%%%%%%%%%%%%%%%%%%%%%%%%
\Big\langle T\mbox{  }\psi&_{L}(X_1)\psi_{R}^{\dagger}(X_2)\Big\rangle \sim 
\frac{(2x_1)^{\gamma_1}(2x_2)^{1+\gamma_2}+(2x_1)^{1+\gamma_2}(2x_2)^{\gamma_1}}{2(x_1-x_2 -v_h \tau_{12})^{S} (-x_1+x_2 -v_h \tau_{12})^{S}} \\
\times&\frac{1}{ (x_1+x_2 -v_h \tau_{12})^{Z} (-x_1-x_2 -v_h \tau_{12})^{Y}(-x_1-x_2 -v_F \tau_{12})^{0.5}}\\
\label{SS}
\end{aligned}
\end{equation}

\small
\begin{bf}Case II : $x_1$ and $x_2$ on opposite sides of the origin\end{bf} \\ \scriptsize

\begin{equation*}
\begin{aligned}
\Big\langle T\mbox{  }\psi&_{R}(X_1)\psi_{R}^{\dagger}(X_2)\Big\rangle \sim 
\frac{(2x_1)^{1+\gamma_2}(2x_2)^{\gamma_1} }{2(x_1-x_2 -v_h \tau_{12})^{A} (-x_1+x_2 -v_h \tau_{12})^{B}} \\
\times&\frac{(x_1+x_2)^{-1}(x_1+x_2 + v_F \tau_{12})^{0.5}}{ (x_1+x_2 -v_h \tau_{12})^{C} (-x_1-x_2 -v_h \tau_{12})^{D} (x_1-x_2 -v_F \tau_{12})^{0.5}}\\
%%%%%%%%%%%%%%%%%%%%%
&\hspace{2cm}+\frac{(2x_1)^{\gamma_1} (2x_2)^{1+\gamma_2}}{2(x_1-x_2 -v_h \tau_{12})^{A} (-x_1+x_2 -v_h \tau_{12})^{B}} \\
\times&\frac{(x_1+x_2)^{-1}(x_1+x_2 - v_F \tau_{12})^{0.5}}{ (x_1+x_2 -v_h \tau_{12})^{D} (-x_1-x_2 -v_h \tau_{12})^{C} (x_1-x_2 -v_F \tau_{12})^{0.5}}\\\\\\
%%%
\Big\langle T\mbox{  }\psi&_{L}(X_1)\psi_{L}^{\dagger}(X_2)\Big\rangle \sim 
\frac{(2x_1)^{1+\gamma_2}(2x_2)^{\gamma_1} }{2(x_1-x_2 -v_h \tau_{12})^{B} (-x_1+x_2 -v_h \tau_{12})^{A}} \\
\times&\frac{(x_1+x_2)^{-1}(x_1+x_2 - v_F \tau_{12})^{0.5}}{ (x_1+x_2 -v_h \tau_{12})^{D} (-x_1-x_2 -v_h \tau_{12})^{C} (-x_1+x_2 -v_F \tau_{12})^{0.5}}\\
%%%%%%%%%%%%%%%%%%%%%
&\hspace{2cm}+\frac{(2x_1)^{\gamma_1} (2x_2)^{1+\gamma_2}}{2(x_1-x_2 -v_h \tau_{12})^{B} (-x_1+x_2 -v_h \tau_{12})^{A}} \\
\times&\frac{(x_1+x_2)^{-1}(x_1+x_2 + v_F \tau_{12})^{0.5}}{ (x_1+x_2 -v_h \tau_{12})^{C} (-x_1-x_2 -v_h \tau_{12})^{D} (-x_1+x_2 -v_F \tau_{12})^{0.5}}\\
\end{aligned}
\end{equation*}

%%%%%%%%%%%%%%%%%%%%%%%%%%%%%%%%%%%%%%%%%%%%%%%
\begin{equation}
\begin{aligned}
%%%%%%%%%%%%%%%%%%%%%%%%%%%%%%%%%%%%%%%%%%%%%%%
\Big\langle T\mbox{  }\psi&_{R}(X_1)\psi_{L}^{\dagger}(X_2)\Big\rangle \sim \mbox{ }0\\
%%%%%%%%%%%%%%%%%%%%%%%%%%%%%%%%%%%%%%%%%%%%%%%
\Big\langle T\mbox{  }\psi&_{L}(X_1)\psi_{R}^{\dagger}(X_2)\Big\rangle \sim  \mbox{ }0\\
\label{OS}
\end{aligned}
\end{equation}
\normalsize
where
\footnotesize
\begin{equation}
Q=\frac{(v_h-v_F)^2}{8 v_h v_F} \mbox{ };\mbox{ }  X=\frac{|R|^2(v_h-v_F)(v_h+v_F)}{8  v_h (v_h-|R|^2 (v_h-v_F))}  \mbox{ };\mbox{ }C=\frac{v_h-v_F}{4v_h}
\label{luttingerexponents}\end{equation}
\normalsize
The other exponents can be expressed in terms of the above exponents.
\footnotesize
\begin{equation*}
\begin{aligned}
&P= \frac{1}{2}+Q  \mbox{ };\hspace{0.8 cm}    S=\frac{Q}{C}( \frac{1}{2}-C)   \mbox{ };\hspace{0.85 cm}      Y=\frac{1}{2}+X-C  ;           \\
& Z=X-C\mbox{ };\hspace{0.8 cm}      A=\frac{1}{2}+Q-X \mbox{ };\hspace{0.8 cm}   B=Q-X  \mbox{ };\hspace{1 cm}   \\
&D=-\frac{1}{2}+C   \mbox{ };\hspace{.6 cm}      \gamma_1=X                \mbox{ };\hspace{1.65 cm}    \gamma_2=-1+X+2C;\\\\\\
\end{aligned}
\end{equation*}
\normalsize

\section*{Funding}
A part of this work was done with financial support from Department of Science and Technology, Govt. of India DST/SERC: SR/S2/CMP/46 2009.\\
%--------------------------------------------------BIBLIOGRAPHY----------------------------------------------------------------------------------------------------

\bibliographystyle{apsrev4-1}
\bibliography{ref}
\normalsize

\end{document}